\documentclass[aps,pre,twocolumn,groupedaddress,showkeys]{revtex4-1}
\usepackage{graphicx}
\usepackage{dcolumn}
\usepackage{bm}
\usepackage{citesort}
\usepackage[percent]{overpic}

\def \const {{\rm const}}
\def \Re {{\rm Re}\, }

\def \be {\begin{equation}}
\def \ee {\end{equation}}
\def \ba {\begin{eqnarray}}
\def \ea {\end{eqnarray}}

\begin{document}

\title{Flow stabilization with active hydrodynamic cloaks}

\author{Yaroslav A. Urzhumov and David R. Smith}
\affiliation{Center for Metamaterials and Integrated Plasmonics, Pratt School of Engineering, Duke University, Durham, North Carolina 27708, USA}
\email{yaroslav.urzhumov@duke.edu}

\date{\today}

\begin{abstract}
We demonstrate that fluid flow cloaking solutions based on active hydrodynamic metamaterials exist for two-dimensional flows past a cylinder in a wide range of Reynolds numbers, up to approximately 200. Within the framework of the classical Brinkman equation for homogenized porous flow, we demonstrate using two different methods that such cloaked flows can be dynamically stable for $Re$ in the range 5-119. The first, highly efficient, method is based on a linearization of the Brinkman-Navier-Stokes equation and finding the eigenfrequencies of the least stable eigen-perturbations; the second method is a direct, numerical integration in the time domain. We show that, by suppressing the Karman vortex street in the weekly turbulent wake, porous flow cloaks can raise the critical Reynolds number up to about 120, or five times greater than for a bare, uncloaked cylinder.

\end{abstract}


\maketitle

Metamaterials (MMs), or artificially structured composites, have been proposed as a supplement to natural, molecular media that extends material properties into the domains not covered by naturally formed substances. Supplemented with a macroscopic design methodology, known as transformation optics (acoustics, and so on), MMs enable a new wave applications by offering extra flexibility in manipulating the dynamics of waves and matter.
Recently, it was proposed~\cite{urzhumov_smith_prl11} that hydrodynamic metamaterials (HDMM) consisting of a fluid-filled solid matrix with a variable fluid permeability can enable exotic flow regimes, in which pressure and velocity differentials are confined to a small volume occupied by the permeable shell. These flows were demonstrated in three dimensions, for a spherical object surrounded by a concentric porous shell with anisotropic, and partially negative, permeability. As explained below, such a property would necessitate the use of active hydrodynamic metamaterials, defined as HDMMs with active elements that can accelerate the fluid; the latter elements can be devised, for example, from electrically powered and controlled micro-pumps~\cite{najafi_patent06,hua_gulari06}.

The wider concept of active, or externally powered, MMs is currently a subject of fundamental studies in the fields of electrical engineering, optics, and acoustics. Here, before attempting to propose a physical implementation of an active HDMM, we wish to address the question whether such media would be of any benefit in hydro- or aerodynamical engineering.
This paper serves to demonstrate that active HDMMs are capable of leveraging the critical Reynolds number ($Re$) at which the flow past an object transitions from laminar to weakly turbulent.


In comparison with the three-dimensional flow past a sphere, the 2D case is substantially more challenging. From the theoretical perspective, exact analytical solutions for flow past a cylinder are not possible even in the limit of small $Re$; Oseen's equation can be used to obtain the lowest-order approximation~\cite{landau_lifshitz_fluid}.
Thus, 2D flows are more sensitive to the approximate treatment (or neglect) of the momentum advection term ($(\vec u \cdot \vec \nabla)\vec u$) in the Navier-Stokes equations. This term is responsible for chaotic motions that are bound to grow from small perturbations when $Re$ exceeds a certain critical value; well-developed stochasticity of a flow with stationary boundary conditions is known as turbulence~\cite{landau_lifshitz_fluid}.

For 2D flows past a cylinder, the growth of perturbations is a concern even at low $Re$ (less than a hundred), where the spontaneous formation and separation of oscillating eddies in the downstream wake gives rise to the Karman vortex street~\cite{park_hendricks94}. Vortex shedding has lots of undesirable practical consequences, including vibrations of long fluid-protruding objects (towers, chimneys, aerials) and bubble generation in gas-saturated liquids (submarine periscopes). Engineering effort was made to control the wake and suppress vortex shedding by manipulating the separation of vortices from the boundary layer by both passive means such as fins and other vortex spoilers~\cite{king77}, as well as active devices~\cite{park_hendricks94,glezer_amitay02,lee_mallinson03}. In this paper, we propose that fluid-permeable active flow ``cloaks" can suppress the formation of eddies in a wide range of $Re$ numbers, thus leveraging the onset of turbulence.

\begin{figure}
\centering
\begin{tabular}{cc}
\begin{overpic}[width=0.4\columnwidth]{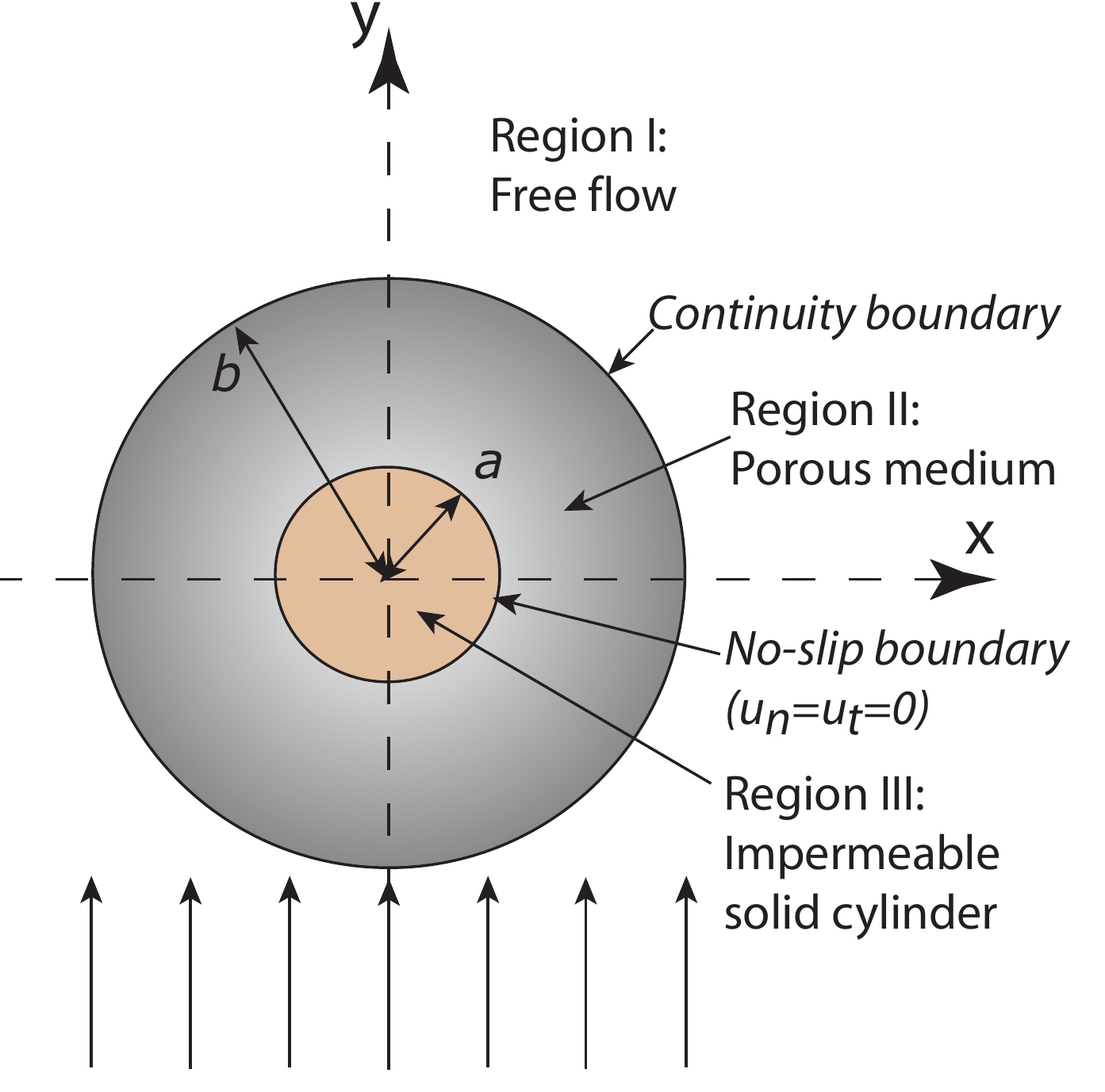}
\put(15,95){(a)}
\end{overpic}
&
\begin{overpic}[width=0.5\columnwidth]{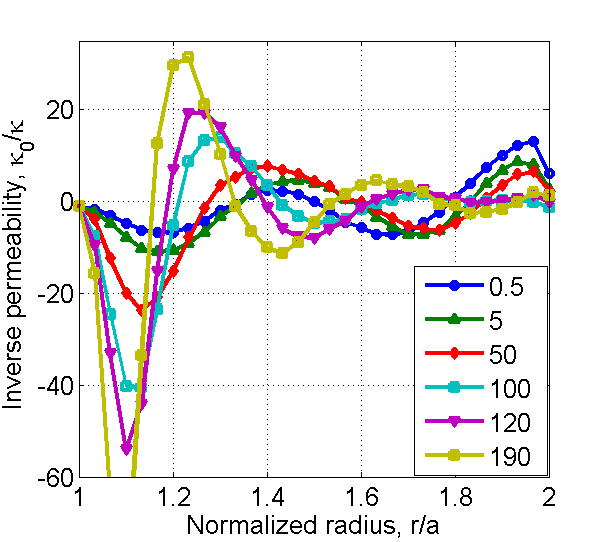}
\put(15,75){(b)}
\end{overpic}
\\
\begin{overpic}[width=0.46\columnwidth]{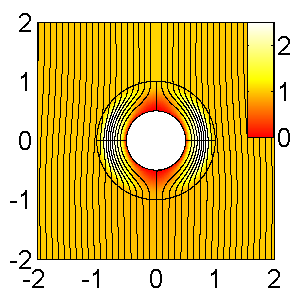}
\put(15,85){(c)}
\end{overpic}
&
\begin{overpic}[width=0.45\columnwidth]{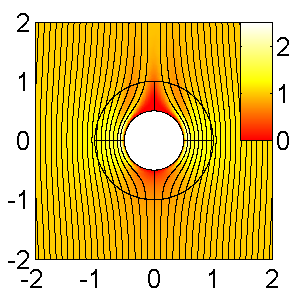}
\put(15,85){(d)}
\end{overpic}
\end{tabular}
\caption{(color online).
(a) Schematic of the structure and the problem setup.
(b) Permeability profiles designed for flows with various flow rates, labeled by the corresponding Reynolds number $Re=\rho u_0 a/\mu$.
Steady-state cloaking solutions:
distribution of the vertical velocity $u_y$ normalized to $u_0$ (shade) and streamlines of flow (black lines) for solutions optimized at various flow rates:
(c) $Re=0.5$,
(d) $Re=190$.
}
\label{fig:steady}
\end{figure}


We begin our analysis by postulating the time-dependent nonlinear Brinkman equations for the macroscopic description of flows~\cite{bear72,koplik_levine83,masliyah_ven87,nield_bejan06} in HDMMs:

\ba
\frac{\partial}{\partial t}\left(\epsilon \rho\right) + \vec \nabla \cdot (\rho \vec u) = 0, \\
\frac{\rho}{\epsilon}\left(\frac{\partial \vec u}{\partial t} + (\vec u \cdot \vec \nabla) \frac{\vec u}{\epsilon} \right)
= - \vec \nabla p - \mu\kappa^{-1}\vec u
\nonumber \\
+\vec\nabla \left[ \frac{\mu}{\epsilon}\left(\overleftarrow{\nabla} \vec u +(\overleftarrow{\nabla} \vec u)^{T}\right)
-\frac{2\mu}{3\epsilon}(\vec\nabla\cdot\vec u)\right],
\label{eqn:Brinkman_full}
\ea
We wish to simplify the analysis by setting the porosity parameter $\epsilon$, which represents the filling fraction of the fluid, to a constant value of unity. This approximation is consistent with the expectation that the desirable porous media will have a very small filling fraction of the solid phase, which would allow them to have a high permeability (low resistance to the flow).
Generalization of the forthcoming analysis to specific values of $\epsilon<1$ is straightforward.

With this simplification, the steady-state form of the Brinkman equation for an incompressible fluid ($\rho=\const$) with a constant free-stream dynamic viscosity $\mu$ and coordinate-dependent inverse permeability tensor $\kappa^{-1}$ becomes
\be
\rho (\vec u \cdot \vec \nabla) \vec u  =
 - \vec \nabla p + \mu \nabla^2 \vec u  - \mu\kappa^{-1}\vec u,
\label{eqn:Brinkman_steady}
\ee
combined with the steady-state continuity equation,
\be
\vec\nabla \cdot \vec u =0.
\label{eq:continuity}
\ee

Within the framework of Eq.~(\ref{eqn:Brinkman_full}), we solve the problem of two-dimensional flow cloaking~\cite{urzhumov_smith_prl11}, or wake elimination.
We then proceed to demonstrating that wake elimination indeed results in flow stabilization at flow rates that normally result in turbulence. The schematic of the geometry under consideration is shown in Fig.~\ref{fig:steady}(a).
The boundary conditions are no-slip ($\vec u=0$) at the surface of an impermeable cylinder ($r=a$), and plug flow in the y-direction ($\vec u = u_0 \hat y$) at infinity ($r\to\infty$). Elsewhere, including the transition from porous to free flow domains at $r=b$, the continuity conditions~\cite{bars_worster06} are assumed for the velocity field and for the fluid stress tensor, including its pressure and shear components. Whenever a specific calculation occurs, we assume $a=0.5$~mm, $b=2a$, $\mu=1\cdot 10^{-3}$~Pa$\cdot$s, and $\rho=1$~g/cm$^3$. When referring to normalized permeability, the normalization constant $\kappa_0=a^2/4$ is assumed.

Previously, approximate solutions were reported for the case of flow past a spherical cloak~\cite{urzhumov_smith_prl11}.
The proposed technique was based on the introduction of the (Stokes) stream function and solving the linear equation in the regime with vanishingly small $Re$. However, for the flow past a cylinder, it is well known that the linear Stokes equation cannot give meaningful solutions that satisfy the boundary conditions at $r=a$ and infinity; the origin of the problem is the existence of a logarithmic fundamental solution. This issue can be partially alleviated by considering Oseen's equation, which uses an approximate, linearized form of the advection term instead of neglecting it entirely. Still, this term, due to its angular symmetry, would break down the separation of variables in the polar coordinates, thus rendering the method of Ref.~\cite{urzhumov_smith_prl11} much more difficult to apply. Thus, we choose to obtain the wake-free (cloaking) solutions numerically by employing a gradient-based optimization technique.

It was shown in Ref.~\cite{urzhumov_smith_prl11} that wake-free solutions exist when the permeability is spherically symmetric but locally anisotropic with the spherically-uniaxial type of anisotropy.
Here, we show that approximately wake-free designs can be obtained by assuming locally {\it isotropic} and cylindrically-symmetric permeability $\kappa(r)$.
For solving the forward steady-state problem with any given $\kappa(r)$,
our approach employs a standard two-dimensional Galerkin (finite element) method provided by simulation software COMSOL Multiphysics~\cite{comsol11}, specifically, in its ``free and porous flow" interface, which includes both Brinkman and Navier-Stokes equations.
The simulation domain throughout this work is a rectangle of width $W=20a$ horizontally and height $H=60a$ vertically.
The inflow boundary is positioned at $y=-H/3$ and the outlet is at $y=2H/3$, with the center of the cylinder at the origin.
The forward problem is then embedded into a gradient-based optimization algorithm (SNOPT), which is also part of the same program.
This integrated forward-inverse solving strategy lets us use the built-in semi-analytic sensitivity analysis, which produces the gradient of the optimization goal with respect to all optimization variables in one step, which is then used by the gradient-assisted optimization routine.
The optimization goal is to minimize the norm of velocity deviation from a perfectly uniform plug flow, i.e.
\be
W = \int |\vec u - \vec u_0|^2 dV,
\label{eq:opt_goal}
\ee
where the integral is taken over the exterior of the cylinder of radius $r=b$ which represents the ``cloak". The quantity $W$ can be seen as a measure of wake in the flow past the structure.
The optimization variables are the nodal values of the unknown permeability function, $\kappa(r)$, discretized on a certain finite element mesh.
These nodal values are then extruded and interpolated onto the entire annulus $a\le r \le b$.

In this fashion, we obtain a fully converged solution with the flow rate corresponding to $Re=0.5$, and then use that solution as an initial guess for the next run of the optimization solver at a slightly higher inflow velocity $u_0$. This procedure yields fully converged (in the sense of the gradient-based optimization) solutions, showing very little velocity perturbation (\ref{eq:opt_goal}) in the exterior of the permeable cylinder. Several such solutions are illustrated in Fig.~\ref{fig:steady}(b-d).

A key feature of the obtained solutions is that they contain regions where $\kappa^{-1}<0$. 
While we are not aware of physical implementations of negative $\kappa$ as {\it passive} porous media, 
we may conceive it to be implemented as an {\it active HDMM}. The latter is defined here as a fluid-permeable structure that applies an accelerating force to the fluid. To understand how an active HDMM can implement the proposed solutions, one should realize that
the Brinkman Eq.~(\ref{eqn:Brinkman_steady}) is nothing more than the momentum transport equation with three volumetric forces: pressure gradient force, viscous shear force and, lastly, the force exerted by the solid components of the HDMM upon the fluid,
\be
\vec F_s = - \mu\kappa^{-1}\vec u.
\label{eq:Brinkman_force}
\ee
In passive media (or media with $\kappa>0$) $\vec F_s$ is typically contra-directed with the average flow direction $\vec u$, so that it does work against the flow ($\vec F_s \cdot \vec u<0$) and thus decelerates it. However, in an HDMM an embedded array of thrust-generating elements (such as mini- or micro-pumps~\cite{najafi_patent06,hua_gulari06}) may generate a force $F_s$ with an arbitrary direction relative to $\vec u$, including $\vec F_s \cdot \vec u>0$. Therefore, our HDMM cloaking solutions are to be implemented in practice as follows.
Once the spatial distributions $\kappa(\vec r)$ and macroscopic fluid velocity $\vec u(\vec r)$ are found using the technique presented here,
negative permeability regions are replaced with an equivalent body force, $\vec f_{thrust}$, using the trivial relationship
$\mu \kappa^{-1}(r) \vec u(\vec r) = \mu \kappa_p^{-1}(r) \vec u(\vec r) - \vec f_{thrust}(\vec r)$, where $\kappa_p(r)>0$ is an arbitrary positive function of radius. We emphasize that the origin of negative permeability in our results is the ansatz of Eq.~(\ref{eq:Brinkman_force}) used to express the volumetric force needed to sustain the desired flow pattern ($\vec F_s$) in terms of the macroscopic fluid velocity ($\vec u$). This ansatz is used merely for historical reasons, and it is neither expected to be valid at high $Re$ nor required to obtain the types of flows reported here.

In order to address the question of dynamic stability of the stationary solutions obtained, we first employ the method of small perturbation~\cite{kawaguti_jpsj55}. The technique is based on the linearization of the full time-dependent Eq.~(\ref{eqn:Brinkman_full}) with respect to the steady-state solution ($\vec u_s$, $p_s$) of the nonlinear Eq.~(\ref{eqn:Brinkman_steady}). Assuming that the velocity and pressure perturbations ($\vec v$, $p'$) depend on time as $e^{-\lambda t}$, the following eigenvalue problem can be formulated:
\be
\rho (\vec u_s \cdot \vec \nabla) \vec v +\rho (\vec v \cdot \vec \nabla) \vec u_s +
  \vec \nabla p' - \mu \nabla^2 \vec v  + \mu\kappa^{-1}\vec v = \lambda \rho \vec v,
\label{eqn:Brinkman_linearized}
\ee
which must be supplemented by the linearized continuity equation, $\vec \nabla \cdot \vec v=0$.

\begin{figure}
\centering
\includegraphics[width=0.9\columnwidth]{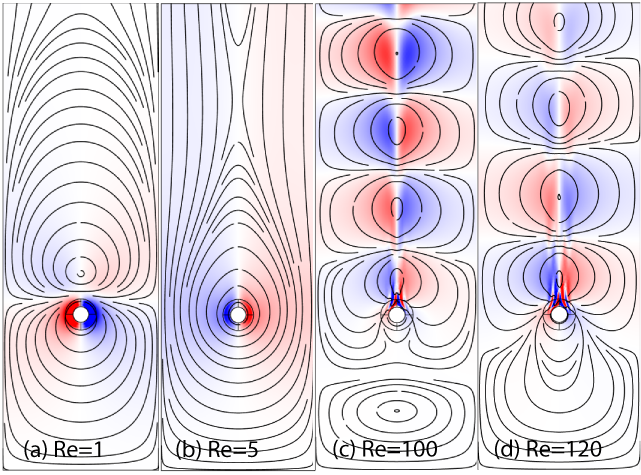}
\caption{(color online).
Stability analysis using harmonic perturbation method for several fluid cloaks.
Distribution of velocity perturbation (shade) and its streamlines (black lines) for the eigenmode with the lowest $Re(\lambda)$; cloaks with (a) $Re=1$ (unstable), (b) $Re=5$ (stable), (c) $Re=100$ (stable), (d) $Re=120$ (unstable).
}
\label{fig:perturbation}
\end{figure}

Once the eigenvalue problem is solved, the eigenvalues $\lambda_n$ can be used to determine whether the solution is stable with respect to small perturbations. The steady-state solution is unconditionally stable when all eigenvalues $\lambda_n$ lie in the right half-plane, i.e. $\Re \lambda_n\ge 0$~$\forall n$. Thus, the critical velocity at which the transition from laminar to turbulent flow occurs can be determined by looking at $\Lambda = \min_n \Re \lambda_n$ as a function of $u_0$; once $\Lambda$ becomes negative, the flow is no longer stable.

First, we calibrated this method by solving the classical problem of laminar flow past a bare, impermeable cylinder of radius $a$ (results not shown). The spectra we obtain indicate that such flow is unconditionally stable for Reynolds numbers $Re=\rho u_0 a/\mu$ up to $Re_{cr}^{(0)}= 23.5$, and bears unstable eigenmode(s) above that Reynolds number. The instability-causing eigenmode in the latter regime has the appearance of the familiar Karman vortex street. This finding is in general agreement with other theoretical and experimental results~\cite{kawaguti_jpsj53,tritton59}.

\begin{table}\centering
\begin{tabular}[t]{|c||c|c|c|}
\hline
 $Re$ & $\lambda_1$ & $\lambda_2$ & $\lambda_3$  \\
 \hline\hline
0.5 & $-4.093$ & $0.3456 \pm 0.0576i$  &  $0.4466 \pm 0.1393i$  \\
1 & $-2.621$ & $0.6843 \pm 0.1914i$  &  $0.818 \pm 0.4684i$  \\
4 & $-0.4757$ & $3.008 \pm 0.7027i$ &  $3.4721 \pm 2.4045i$  \\
\hline
5 & $ 0.3651$ & $3.823 \pm 0.5922i$ & $4.477 \pm 3.008i  $  \\
10 & $3.369$ & $ 7.4493$ & $8.7070$  \\
50 & $18.618 \pm 40.715i$ & $ 48.461 \pm 55.731i$ & $51.655$  \\
100 & $9.748 \pm 125.92i$ & $ 93.835$ & $ 94.886 \pm 96.484i$  \\
119 & $8.950 \pm 154.387i$ & $ 113.795 \pm 114.267i$ & $ 114.164$  \\
\hline
120 & $-11.718 \pm 134.177i$ & $ 115.740$ & $ 118.602 \pm 73.088i$  \\
190 & $-21.490 \pm 211.938i$ & $ 85.055$ & $ 182.137$  \\
 \hline
\end{tabular}
\vskip2mm
\caption{The three lowest eigenvalues (ordered by increasing $\Re \lambda$)
for cloaks designed for different $Re$.}
\label{table:eigenvalues}
\end{table}

We then apply this method to porous cloaks with various flow rates, which are labeled by the corresponding Reynolds number of the flow, $Re = \rho u_0 a/\mu$, calculated with respect to the ``cloaked" cylinder radius $a$.
The lowest three eigenvalues for cloaks with various $Re$ are listed in Table~\ref{table:eigenvalues}. Noticeably, an unstable eigenmode exists in the range $Re\le 4$, which becomes stable at $Re\approx 5$ and remains stable at higher $Re$. A snapshot of this eigenmode's velocity perturbation profile is shown in Fig.~\ref{fig:perturbation}(a,b), both below (a) and above (b) the transition $Re$. This behavior --- flow stabilization with increasing flow rate --- is drastically different from the usual laminar-turbulent flow transitions.
Apparently, this form of instability can be attributed to the presence of an active (negative-permeability) medium, which adds momentum to the flow in proportion to the local velocity (see Eq.~(\ref{eq:Brinkman_force}) and the discussion below it). Not surprisingly, a hypothetical medium with a momentum pump whose strength grows with the flow rate acts as an amplifier of local fluctuations, causing the system to exhibit an instability even at very small $Re$ well within the Stokes flow regime.

In the range $Re=5-119$ (see Table~\ref{table:eigenvalues} and Fig.~\ref{fig:perturbation}(c)), all eigenmodes existing in our simulation domain are stable, which leads us to believe that the steady-state solutions obtained in this regime are unconditionally stable. This conclusion is further confirmed below by another method. At $Re=120$, an eigenmode corresponding to the formation of an oscillating Karman vortex street past the cylinder (Fig.~\ref{fig:perturbation}(d)) becomes unstable, and remains unstable for all higher $Re$ that were simulated (up to $Re=192$). One may observe from Fig.~\ref{fig:perturbation}(d) that the eigenmode leading to flow instability in that regime appears to be seeded by the perturbation in the small region past the cylinder near its back surface, where the velocity deviation (or wake) is not completely eliminated in the stationary solution (see Fig.~\ref{fig:steady}(d)).
Since the cloaking structures obtained through numerical optimization of the permeability deviate noticeably from a perfect wake-free goal (\ref{eq:opt_goal}) at $Re>100$, one can therefore expect that a more accurate cloak with a much smaller value of (\ref{eq:opt_goal}) would also possess a more stable perturbation spectrum.

\begin{figure}
\centering
\begin{tabular}{ccc}
\includegraphics[width=0.34\columnwidth]{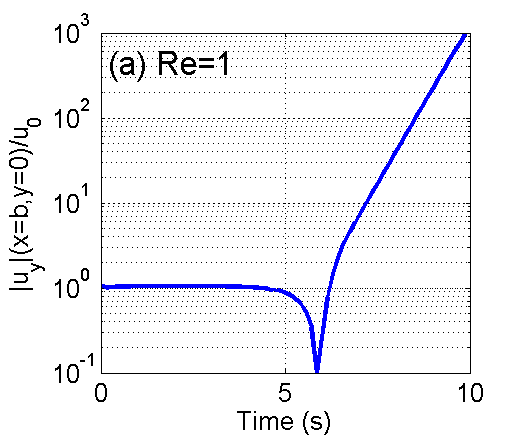}&
\includegraphics[width=0.275\columnwidth]{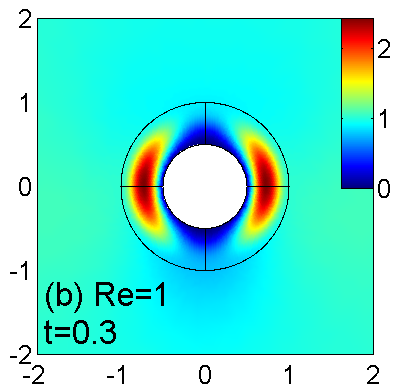}&
\includegraphics[width=0.29\columnwidth]{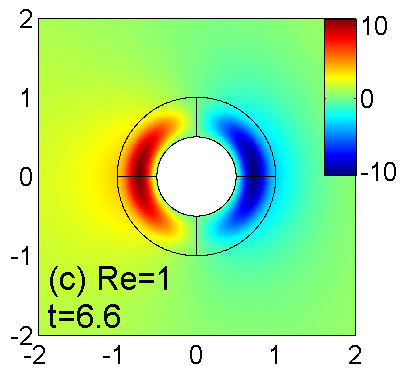}\\
\includegraphics[width=0.34\columnwidth]{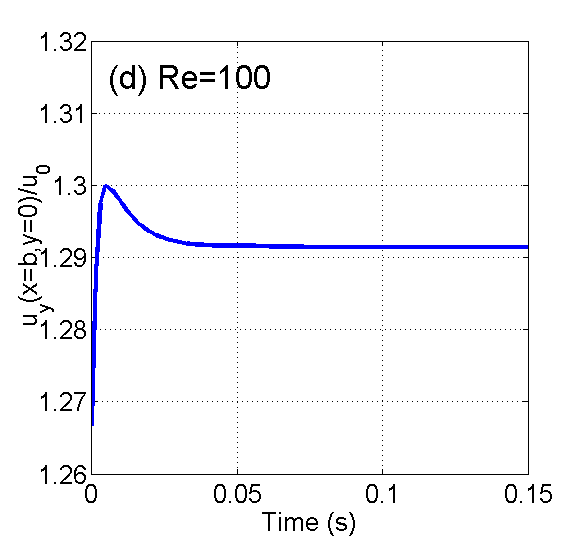}&
\includegraphics[width=0.28\columnwidth]{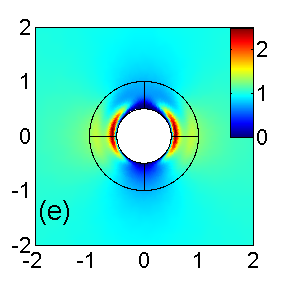}&
\includegraphics[width=0.281\columnwidth]{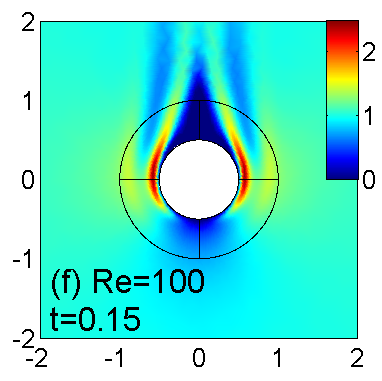}\\
\end{tabular}
\caption{(color online).
Time-dependent analysis for fluid cloaks optimized at different $Re$.
Cloak operating at $Re=1$: (a) vertical velocity on the surface of the cloak at position $x=b, y=0$ vs time (log scale);
(b) $v_y$ velocity component profile at $t=0.3$~s (before instability) and (c) $t=6.6$~s (after unstable eigenmode growth).
At $Re=100$: (d) vertical velocity at a surface point vs time;
(e) $v_y$ snapshot at $t=0.0015$~s (before stabilization) and (f) $t=0.15$~s (after stabilization).
}
\label{fig:transient}
\end{figure}

To validate the findings of our perturbative stability analysis, we perform full transient simulations based on the full time-dependent Eq.~(\ref{eqn:Brinkman_full}). The initial value problem was solved for cloaks with several $u_0$, with the same boundary conditions as were used for the stationary analysis, with the initial values $\vec u(x,y,t=0) = u_0 \hat y$ and $p(x,y,t=0)=0$.

For the cloak designed to operate at $Re=1$, which, for the cylinder dimension $2a=1$~mm corresponds to inflow velocity $u_0=2$~mm/s,
we observe that the solution quickly (within 0.1-0.2~s) reaches the steady-state solution that was found in the stationary analysis (Fig.~\ref{fig:transient}(a-b)). However, as evident from Fig.~\ref{fig:transient}(a), at $t\approx 5$~s since the simulation start, the velocity in the vicinity of the structure begins to grow exponentially, and at $t=6.6$~s (Fig.~\ref{fig:transient}(c)) the velocity profile matches precisely with the unstable eigenmode predicted by the perturbation analysis (see Fig.~\ref{fig:perturbation}(a)). Pure exponential growth of this eigenmode continues throughout the end of the simulation period ($10$~s).

Similar behavior, but on a longer temporal scale is observed for the cloak designed to operate at $Re=4$ (not shown). As seen from Table~\ref{table:eigenvalues}, the unstable eigenmode is characterized by the negative real part $\Re \lambda = -0.4757$, which is less by factor $0.18$ than $\Re \lambda = -2.62$ of this eigenmode at $Re=1$. Thus, one may expect that the time needed for the instability to develop is about $1/0.18\approx 5$ times greater than in the previous case. This prediction is confirmed by the transient simulation: the instability begins to grow at $t\approx25$~s (not shown).

The cloak designed for $Re=100$ (inflow velocity $u_0=0.2$ m/s) shows a drastically different behavior. The flow reaches steady state in about 0.04~s since initialization time, and remains completely unchanged throughout the rest of the simulation period $T=H/u_0=0.15$~s,
which is chosen such that the fluid starting at the inlet has enough time to travel throughout the very long simulation domain. This confirms that the steady-state solution found above is dynamically stable. Virtually identical behavior is observed in the flow with $Re=119$ (not shown), which is the upper limit of the dynamically stable regime in our calculations.

\begin{figure}
\centering
\begin{tabular}{ccc}
\begin{overpic}[width=0.49\columnwidth]{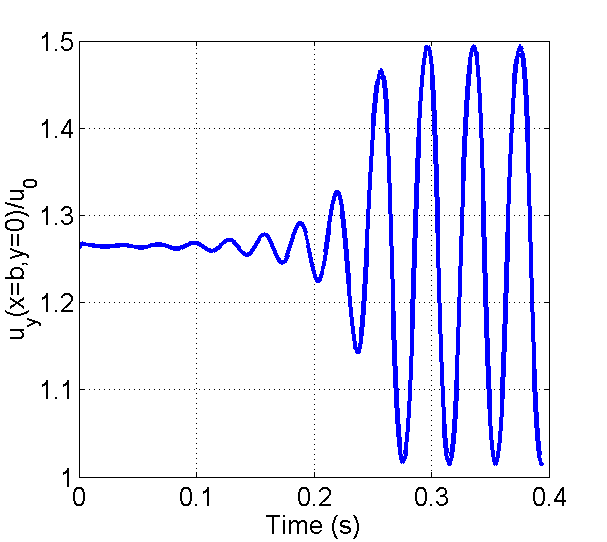}
\put(15,70){(a)}
\end{overpic}
&
\begin{overpic}[width=0.25\columnwidth]{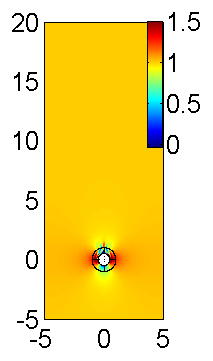}
\put(15,80){(b)}
\end{overpic}
&
\begin{overpic}[width=0.25\columnwidth]{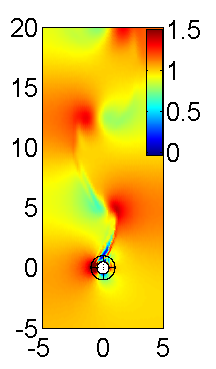}
\put(15,80){(c)}
\end{overpic}
\end{tabular}
\caption{(color online).
Transition to turbulence in the designed cloaks.
At $Re=190$: (a) vertical velocity at a surface point $(x=b,y=0)$ vs time;
(b) $v_y$ profile at $t=0.004$~s (before instability growth) and (b) $t=0.4$~s (after considerable growth of instability).
}
\label{fig:turbulent}
\end{figure}

Above the transition ($Re=120$), numerical integration of time-dependent equations reveals an instability (Fig.~\ref{fig:turbulent}).
The spatial profile of velocity disturbance, and its oscillating nature allows one to immediately identify the vortex sheet eigenmode shown in Fig.~\ref{fig:perturbation}(d). This behavior is seen in the entire range of velocities studied ($Re=120-192$). The growth rate of this instability during its unsaturated (exponential) growth, as well as its oscillation frequency are consistent with correspondingly the real and the imaginary parts of the lowest eigenvalue ($\lambda_1$) shown in Table~\ref{table:eigenvalues}.

The close agreement between the two independent stability analysis methods is evidence, that, within the applicability domain of the Brinkman-Navier-Stokes equations, the fluid-permeable structure described above is able to increase the critical velocity $u_{cr}$, which marks the transition from laminar to turbulent flow, by factor of five. According to the law of hydrodynamic similarity, $u_{cr}$ is a function of the object diameter, $2R$ (which also equals its cross-section in two dimensions), kinematic viscosity $\nu=\mu/\rho$, and a single dimensionless coefficient, $Re_{cr}$ which itself is a function of only the object shape:
$u_{cr} = \frac{Re_{cr} \nu}{R}$.
Thus, the increase in $u_{cr}$ can be stated
either as an increase in $Re_{cr}$, the critical Reynolds number, or
as a reduction of effective diameter, $2R_{\rm eff}$, defined according to $u_{cr} = \frac{Re_{cr}^{(0)}\nu}{ R_{\rm eff}}$, where $Re_{cr}^{(0)}=23.5$ is the critical $Re$ of the bare cylinder. In that sense, the fluid sees the cylinder of radius $R$ surrounded by the cloaking structure as a cylinder of radius $R_{\rm eff}$, which can be substantially smaller than $R$, according to the findings above.

In conclusion, we have demonstrated that approximate fluid flow cloaking solutions exist for flows past a cylindrical obstacle.
These solutions require a mixture of positive and negative permeability inside a cylindrical shell, which can be locally isotropic.
We show that negative permeability gives rise to dynamic instability for flows with Reynolds numbers $Re<5$. However, a family of solutions exist for all $Re$ in the range $5-119$ where the steady-state flow is unconditionally stable.
This finding hints at the possibility of maintaining laminar flows around cylinders at velocities five times the critical velocity for a bare cylinder. Further minimization of the wake at high Reynolds numbers should lead to dynamically stable solutions at the correspondingly higher velocities.

This work was supported by the NAVAIR division of U.S. Navy through a subcontract with SensorMetrix (Contract No. N68335-11-C-0011).
The authors acknowledge useful conversations with Abraham Varghese (U.S. Navy), Earl Dowell, Kenneth Hall and Donald Bliss (Duke University).

\bibliographystyle{unsrt}
\bibliography{CFD_cloak}

\end{document}